# SINGLE/FEW BUNCH SPACE CHARGE EFFECTS AT 8 GEV IN THE FERMILAB MAIN INJECTOR

D. J. Scott, D. Capista, I. Kourbanis, K. Seiya, M.-J. Yang, Fermilab, Batavia, IL 60510, U.S.A.

*Abstract*

For Project X, it is planned to inject a beam of 3 $10^{11}$ particles per bunch into the Main Injector. Therefore, at 8 GeV, there will be increased space charge tune shifts and an increased incoherent tune spread. In preparation for these higher intensity bunches exploratory studies have commenced looking at the transmission of different intensity bunches at different tunes. An experiment is described with results for bunch intensities between 20 and 300 $10^9$ particles. To achieve the highest intensity bunches coalescing at 8 GeV is required, resulting in a longer bunch length. Comparisons show that similar transmission curves are obtained when the intensity and bunch length have increased by similar factors. This indicates the incoherent tune shifts are similar, as expected from theory. The results of these experiments will be used in conjugation with simulations to further study high intensity bunches in the Main Injector.

## INTRODUCTION

A multi-MW proton facility has been established as a critical need for the High Energy Physics Programme of the USA by High Energy Physics Advisory Panel. Project X proposes to use the Fermilab Main Injector synchrotron (MI) as a high intensity proton source capable of delivering 2 MW beam power. This will require 3 times the bunch intensity of current operations. Instabilities associated with beam loading and electron cloud effects are common issues for high intensity proton machines. At the injection energy of 6-8 GeV space charge effects will also be significant and need to be studied to help guide the design and fabrication of the RF cavities and space-charge mitigation devices required for 2 MW operations. The MI intensities for current operations and Project X are listed in Table 1. Bunches with the intensities required for Project X have been created using a coalescing technique [1]. Typically the space charge forces induce coherent and incoherent tune shifts that blow the beam up and reduce the lifetime. Exploratory studies of the effects of incoherent tune shifts are presented.

Table 1: MI Beam Powers.

|  | Current operation | Project X |
|---|---|---|
| Beam power | 400 W | 2000 W |
| Total intensity | $4.0 \times 10^{13}$ | $1.6 \times 10^{14}$ |
| # of Bunches | 492 | 548 |
| Bunch Intensity | $1.0 \times 10^{11}$ | $3.0 \times 10^{11}$ |
| MI cycle rate | 2.2 s | 1.4 s |

## INCOHERENT SPACE CHARGE TUNE SHIFTS

A good introduction to space charge tune shifts can be found here [2]. For a bunched beam of protons with line charge density, λ, the incoherent tune shift in the x and y planes, $\Delta Q_{in,x,y}$ is proportional to:

$$\Delta Q_{in,x,y} \propto \frac{N}{\beta^2 \gamma^3} \frac{F_{x,y} G_{x,y}}{B_f}$$

where $N$ is the number of particles, $\beta$ and $\gamma$ are the relativistic factors, $F_{x,y}$ are derived from the Laslett image coefficients for incoherent tunes shifts, $G_{x,y}$ is a form factor derived from the transverse particle distribution and $B_f$ is the bunching factor or ratio of mean to peak line charge density:

$$B_f = \frac{\langle \lambda \rangle}{\lambda_{peak}} \qquad (1)$$

typically $F_{x,y} < 1, 1 < G_{x,y} < 2$ and $B_f < 1$. From the above expressions it can be seen that as the beam energy increases above ~10 GeV the incoherent tune shifts become small, as expected with space charge effects. It is difficult to measure incoherent tune shifts, typical diagnostics such as BPMs record the coherent, or average, motion of the beam. One method that can be compared with simulation is to look at the beam lifetime, or transmission, the ratio of final to initial beam intensity [3]. Here it is expected that as the incoherent tune shifts increase, i.e. for a higher intensity beam, the beam lifetime will be reduced. The lifetime has been measured for different tunes and different intensity bunches. Qualitative statements can then be made to compare with simulations. Also, from Equation 1, it can be seen that if the bunch length increases in proportion to the bunch intensity then, all other factors being equal, the incoherent tune shift, and hence transmission should be equal. This has been tested and found to be the case.

## MEASUREMENTS

In order to measure the transmission and tune for different intensity bunches with different bunch lengths coalesced and uncoalesced beams were used. The coalesced bunches provided up to 3 $10^{11}$ particles [1], the uncoalesced bunches have 0.5 $10^{11}$ particles. Also, the coalescing process results in a completely filled bucket giving increased bunch lengths compared to the un-coalesced bunches from the booster.

*Tune measurements*

The tune was measured by recording turn by turn BPM (TBT-BPM) data and taking the Fourier transform (FT) of the signal [4]. As there are only a few bunches in the machine the signal from individual BPMs is small, however, by combining the data from all the BPMS in the MI, approximately 100 in both the horizontal and vertical directions, a clear tune signal could be determined. An example of the small signal from each BPM being combined into a clear signal is shown in Figure 1.

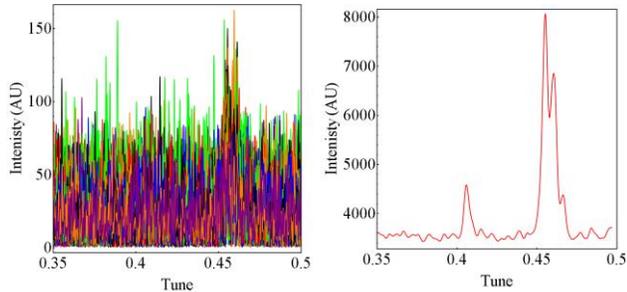

Figure 1: (Left) FT of 100 vertical BPM signals, (right) signals combined.

### Wall Current Monitor Data

Wall Current Monitor (WCM) data, recording the longitudinal beam intensity in the ring, were taken over a number of seconds. For the uncoalesced beam from the booster, 5 main bunches were used, and for the coalesced beam there is one central, high intensity, bunch and two satellite bunches. This is due to the coalescing process not being 100% efficient. Example WCM signals for coalesced and uncoalesced beams can be seen in Figure 2.

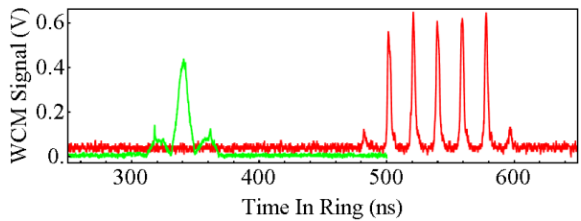

Figure 2: Coalesced and un-coalesced (red) WCM signal.

## WCM DATA ANALYSIS

To calculate the bunch intensities the integral of the WCM data must be taken and then normalised to the beam current in the machine. The decrease in bunch intensity for each frame of WCM data can be calculated to determine the transmission. The bunch lengths can also be easily found from the WCM data.

### Background subtraction

The background signal must be effectively removed from the WCM data. This was done by taking WCM data where there is no beam present and fitting a Gaussian to a histogram of the data points. The peak of the Gaussian fit is assumed to be the WCM background level.

### Normalisation to # of Particles

After background subtraction the integrated WCM signal is equal to the number of particles in the machine. For each experimental shift sets of data for known increases in beam current were taken, measured using the *IBeam* parameter in the MI control system. The integrated WCM signal data was then compared to the *IBeam* data and a line of best fit used to determine the number of particles in a bunch. An example normalisation curve is shown in Figure 3. This data was taken each time the experiment was run as the WCM signal depends on external factors, such as the temperature in the accelerator tunnel.

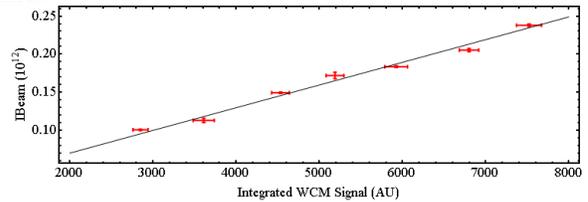

Figure 3: Integrated WCM signal vs IBeam reading.

### Calculating Bunch Lifetime

From the WCM data individual bunches can easily be resolved (see Figure 2). The WCM data spans about 3 seconds and so the transmission rates for bunches can easily be determined. Some example of the bunch intensity decreasing over time can be seen in Figure 4.

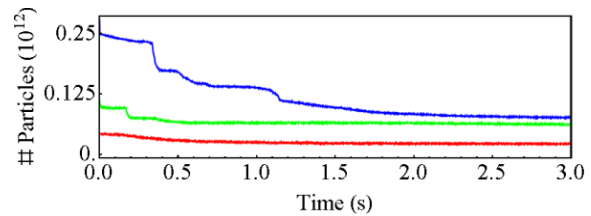

Figure 4: Example bunch intensities vs time.

### Bunch Lengths

The coalesced and un-coalesced bunches have different bunch lengths, which effect the incoherent tune shift. These have been calculated by fitting Gaussians. Figure 5 shows some examples bunches of different intensity.

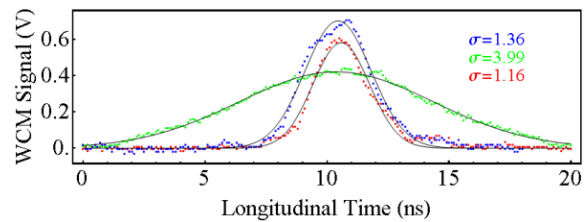

Figure 5: WCM signal for 3 bunches with Gaussian fits.

## TRANSMISSION RESULTS

The transmission rates are shown in Figure 6 for different intensity bunches. Whilst the horizontal or vertical tunes were changed the other tune was held at their nominal value, $Q_h$ =0.47 and $Q_v$ = 0.43. Figure 6 also shows a comparison of the transmission for two bunches with different intensities ($I_b$) and bunch lengths ($\sigma_b$) such that their ratio for Equation 1 is the same. As expected they have very similar transmission curves.

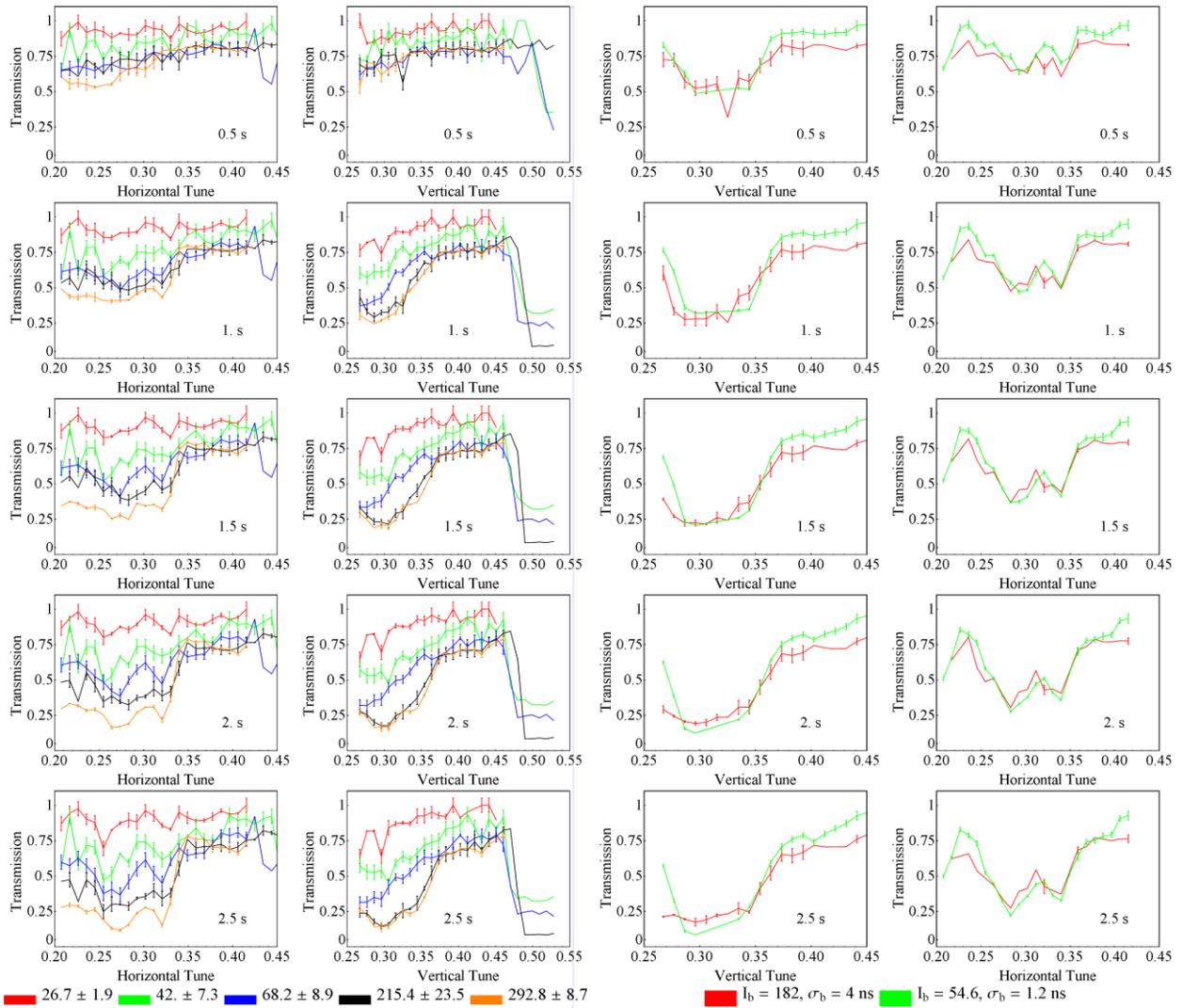

Figure 6: (Left hand side) Transmission after 0.5 to 2.5 seconds for 5 different intensity bunches (red, green, blue, black orange, $10^9$ particles) whilst changing the horizontal and vertical tunes. (Right hand side) Transmission at different tunes after 0.5 to 2.5 seconds of a long high intensity bunch, ($I_b$=182 $10^9$ particles, $\sigma_b$=4 ns) and a short low intensity bunch a ($I_b$=54.6 $10^9$ particles, $\sigma_b$=1.2 ns) giving the same $B_f$.

## CONCLUSIONS

Initial beam experiments in the MI of bunches with intensities similar to those planned for Project X have commenced. These first experiments have looked at the transmission rates of different intensity bunches. These lifetimes should depend upon space charge tune shifts.

As well as confirming that higher intensity bunches have shorter lifetimes qualitative statements can be made from the LHS of Figure 6. For example, the two highest intensity bunches have similar transmission for vertical tune changes and different transmission for horizontal changes. These statements can be checked with simulations to see if similar results can be obtained. The similarity between the transmission rates for the different bunches in the RHS of Figure 6 indicate that the incoherent tune shifts are similar, as expected from theory. The similarity was not as good for lower intensity bunches, and this was attributed to the un-coalesced low intensity beam not being well matched at injection. Further work could study the coherent tune shifts of the high intensity bunches, a good match between theory and experiment has been found for uncoalesced bunches [4].